\documentclass[useAMS,usenatbib]{mn2e}
\usepackage{graphics}
\usepackage{graphicx}
\usepackage{multirow}
\usepackage{amssymb}
\usepackage{color}
\newcommand{\be}{\begin{equation}}
\newcommand{\ee}{\end{equation}}
\newcommand{\bc}{\begin{center}}
\newcommand{\ec}{\end{center}}

\def \LCDM{$\Lambda$CDM}
\def \LCDMu {$\Lambda$CDMu}
\def \HRLCDM {$\Lambda$CDMhr}
\def \LCDMhr {$\Lambda$CDMhr}

\def \sigh{$\sigma_H$}
\def \singles{\textit{Singles}}
\def \pairs{\textit{Pairs}}
\newcommand{\hmsun}{{\,\rm h^{-1}M}_\odot}
\newcommand{\hmpc}{{\,\rm h^{-1}Mpc}}\newcommand{\mpc}{{\,\rm Mpc}}
\newcommand{\hkpc}{{\,\rm h^{-1}kpc}}
\newcommand{\kms}{{\,\rm km \ s^{-1}}}
\def \sig {$\sigma_H$}

\newcommand\aap{A\&A}%
\newcommand\apj{ApJ}%
\newcommand\mnras{MNRAS}%
\newcommand\aj{AJ}%
\newcommand\apjs{ApJS}%

\begin{document}

 \bibliographystyle{mn2e}

\title{Constrained simulations of the local universe:  II. The nature of the local Hubble flow}
\author[Martinez-Vaquero et al.]
{\parbox[t]\textwidth{Luis A. Martinez-Vaquero$^1$, Gustavo Yepes$^1$, Yehuda Hoffman$^2$, Stefan  Gottl\"ober$^3$, Mira Sivan$^2$}
\vspace*{6pt} \\
$^1$Grupo de Astrof\'{\i}sica, 
Universidad Aut\'onoma de Madrid,
Madrid E-280049, Spain 
\\
$^2$Racah Institute of Physics, 
Hebrew University, 
Jerusalem 91904, Israel 
\\
$^3$ Astrophysikalisches Institut Potsdam, An der Sternwarte 16,
            14482 Potsdam, Germany
}
\date{\today}

\maketitle

\begin{abstract}

Using a suite of N-body simulations in different Cold Dark Matter (CDM)
scenarios, with cosmological constant (\LCDM ) and without (OCDM, SCDM),  we 
 study the Hubble flow (\sigh) in Local Volumes
(LV) around Local Group (LG) like objects found in these simulations,
and compare the numerical results with the most recent  observations. 
We show  that \LCDM\ and OCDM models exhibit 
 the same behavior of \sigh. Hence, we demonstrate that  the observed
coldness of the Hubble flow is not likely to be a manifestation of the dark
energy, contrary to previous claims. 
The coldness does not constitute a problem by itself but it
poses a problem to the standard \LCDM\ model only if the mean density
within the Local Volume is greater than twice the mean matter cosmic
density. The lack of blueshifted galaxies in the LV, outside of the LG
can be considered as another manifestation of the coldness of the
flow. Finally, we show  that the main dynamical parameter 
that affects the coldness of the flow is the relative 
isolation of the LG, and the absence of nearby Milky Way like objects 
within a distance of about $3\mpc$.

\end{abstract}


\section{Introduction}
\label{sec:intro}

The neighborhood of the Local Group (LG) is often described as being
{\it cold}. This attribute implies that the dispersion of the radial
velocities of galaxies from a pure Hubble flow is small, and the
'smallness' amounts to less than  $100\kms$.  \cite{san72} studied
local departures from a uniform Hubble flow and could only put upper
limits on such departures. This led 
 \citeauthor{san72} 
to conclude that $q_0 \simeq 0$
(where $q_0$ is the deceleration parameter). Later on \cite{san75}
estimated that the upper limit to the mean random motion of field
galaxies is $\lesssim 50 \kms$. These early findings  of Sandage, Tammann 
and their collaborators have been corroborated  and  vigorously
improved  by many others.  \cite{kar03} estimated the radial peculiar velocity
dispersion of all galaxies within $5.5 \mpc$ to be $85 \kms$. 
 This value drops down to $41 \kms$ if members of galaxy groups are
 removed and distance erros are taken into account.   
Using a newer set of Karachentsev's data \cite{tik08} 
found a velocity dispersion of  $97 \kms$,  within $ 7
\mpc$,  which reduces after correction for apex motion and distance errors only slightly to $84 \kms$.  
 \cite{mac05}  compiled the data from three different sources: The Cepheid-based distance measurements of the HST Key Project \citep{fre01}, distance estimates based on the surface brightness fluctuations method \citep{sbf} and  Tully-Fisher distances \citep{tul92}. They 
 fitted the data by $\sigma_H = 88 \pm 20 \kms \times (R /  7 \mathrm{Mpc})$, where  \sigh\ is a measure of the dispersion of the radial velocities around a pure Hubble flow 
 of galaxies within a sphere of radius $R$ 
 (a thorough discussion of the various estimates of \sig\ is given below).

The observational evidences for a  
local cold Hubble flow 
seem to be indisputable.   Yet, the question arises as to why  a \sigh\ of
the order of a few tens of $\kms$ is labeled as 'cold'. Namely, by what
standard   is  it   cold?.  Rich clusters of galaxies provide the first and the most robust evidence for a departure from a pure Hubble flow, with a dispersion of peculiar velocities   of up to $\approx 10^3 \kms$. Compared with the  rich clusters the neighborhood of the LG is definitely cold. 
A statistical   estimate is also  given by the pair weighted velocity dispersion ($\sigma_{12}$) which was measured from the CfA redshift survey to be $\sigma_{12}(r = 1 \hmpc) = 340 \pm 40 \kms$ (Davis \& Peebles 1983).
Another more robust  measure of the  deviation from the Hubble flow is provided by the $\sigma_1$ statistics which measures the one-dimensional rms peculiar-velocity dispersion of galaxies relative to their neighbors within a projected radius of $2\hmpc$ (Davis, Miller \& White 1997). These authors found $\sigma_1=95 \pm 16 \kms$ (for the IRAS survey) and $130 \pm 15 \kms$ for the UGC catalog. The measured $\sigma_1$ is indeed much 'hotter' than the 
$\sigma_H=  25  \kms$ within $R=3.0 \mpc$ \citep{kar09}.
So, with regard to the other measure of the dispersion of peculiar
velocities the immediate neighborhood of the LG is indeed very
cold. However, one should recall that the $\sigma_1$ and $\sigma_{12}$ measures
consider  all galaxies in a given survey. The \sigh\  considered here,
on the other hand,  refers to one particular object, namely the LG,
that resides in a dynamically quiet  environment. It should be
emphasized that the LG is not an atypical object, yet the LG
environment is not  a representative one for Milky Way like objects. 
This implies that, when  addressing the issue of 'coldness' by means of simulations, the selection of the LG-like objects should be carefully done.

The standard model of cosmology 
consists  of a flat Friedmann universe whose mass energy is dominated
by a cosmological constant ($\Lambda$) and cold dark matter (labeled as
\LCDM). It has been recently stated that the cosmological constant, or
its generalization the Dark Energy (DE), manifests itself in the dynamics of the very local universe (\citeauthor{bar01} \citeyear{bar01}, \citeauthor{che04} \citeyear{che04}, \citeyear{che07}). 
These authors argued that the coldness of the local flow is a
manifestation of the existence of the dark energy. This has been
supported by \cite{mac05} who analysed a set of N-body simulations and
concluded that indeed 
\textit{''.. [their] results provide new,
  independent evidence for the presence of dark energy on scales of a
  few  megaparsecs''}. These results,  if correct, would have provided
an independent corroboration  to  the DE component   whose existence is
otherwise  inferred    from observations of distant
 objects and on very large scales of the  Universe.
The  claims of \cite{che07}  have been challenged by 
\cite{hf08}, who 
analyzed a suit of constrained and unconstrained CDM
simulations, identified LG-like objects and studied the flow field
around these. The main result of \cite{hf08} invalidates the model
advocated by \cite{che07} and its basic predictions. The claims  made
by \cite{mac05} have  remained unchallenged till present. 

There are two main issues this paper addresses. 
The  claim that the coldness of the local flow is induced by the dark
energy, and the fact that it is  strongly  contradicted by    
\cite{hf08},
motivated  us  to revisit the problem. 
Then, there is the general problem of the nature of
 the local flow. In particular the question of how  the various
  characteristics of the local neighborhood are affecting the local flow is addressed here.

  The present paper adopts the methodology used by \cite{mac05} and
  later on by \cite{hf08}. Namely, a set of numerical simulations is
  performed, ensembles of LG-like objects are constructed and \sigh\ is
  calculated around these mock objects in the same way as the Hubble
  flow around the actual LG is analyzed. This is further extended here.
  The cosmological models studied here are the presently accepted model
  of cosmology with a cosmological constant, \LCDM, the open CDM (OCDM)
  and the outdated Einstein-De Siter SCDM models (where S stands for
  the standard of the years of the 1990s and CDM stands for cold dark
  matter).  Most of the simulations studied here are constrained and
  thereby are constructed to reproduce within the simulation box the
  observed Large Scale Structure (LSS) on scales larger than $\approx 5
  \hmpc$.  A thorough study of the nature of the local Hubble flow is
  done by studying the dependence of \sigh\ on a multitude of factors,
  including the cosmological parameters, the mass resolution, the very
  local environment and the structure on larger scales. The control
  over the very small scales is achieved by varying the selection rules
  for the LG-like objects and the control over the large scale
  environment is done by performing the constrained simulations. A
  somewhat different selection of mock objects is obtained by
  constraining them to have no nearby blueshifted galaxies.

The structure of the paper is as follows. 
A summary of the CNG data and examination of the different ways of
calculating \sigh\  are given in \S\ref{sec:obs}. 
\S\ref{sec:sim} describes the N-body simulations analyzed here and
 a description of the selection of the LG-like objects is presented in
 \S\ref{sec:sel}.  The analysis of \sigh\ is conducted first along the lines of 
 \cite{mac05} (\S\ref{sec:result}). Then   the dependence of \sigh\ on
 the various  characteristics of the LG is shown in \S\ref{sec:relax}
 and the possible role of  mass resolution, in    \S\ref{sec:res}. 
The lack of nearby blueshifted galaxies as a  manifestation of the coldness of the local flow is analyzed  in \S\ref{sec:blue}.
The paper concludes with a  summary and a general discussion (\S\ref{sec:disc}).

\section{Observations}
\label{sec:obs}

The Catalog of Neighboring Galaxies \cite[][CNG]{kar04} provides the latest and  most comprehensive survey of the velocities of nearby galaxies.
The catalog includes the  distances and radial
velocities of more than 400  galaxies, roughly 300  of which are in the Local Volume
(LV), defined by a sphere of $R=7 \mpc$ about the LG. \cite{tik08} have recently compiled the catalog and calculated \sigh\ of all the CNG galaxies within the LV. Distances are typically measured with $10\%$ errors. 
  \citeauthor{tik08} studied the recent, yet unpublished version of the
  catalog. That  compilation of the data is taken here as the
  observational reference value, against which our numerical 
 results are to be compared.

The local flow is studied here  within  the framework defined by
\cite{tik08}. Their  study ignores non-linear motions within the LG and
therefore all galaxies closer than $1.0 \mpc$ are excluded from the
analysis.  The flow is analyzed by means of measuring the dispersion of
the radial velocities about the Hubble flow of all   galaxies found
within the LV, which is redefined to correspond to the shell of $(1 \ -
\ 7)\mpc$. The choice to consider all galaxies regardless of their
possible membership in  galaxy  groups is driven by the principle of
keeping the selection to be as clear and simple as possible. This
avoids the issue of defining a galaxy group catalog and simplifies the
comparison of observations and simulations.  

There are different ways of defining the scatter  of the line-of-sight
(LOS) velocities from a pure Hubble, \sigh. Consider a catalog of
galaxies whose distances and radial velocities are measured, $\{
(r_i,v_i)\}_{i=1,..., N}$.  
\cite{tik08} estimated the   apex motion of the observer, namely the
LG, with respect to the sample of galaxies in the catalog and
subtracted the apex velocity from the observed LOS  velocities. The
dispersion \sigh\  is then calculated as the  root mean square (RMS)
with respect to the Hubble flow, where the global value of the Hubble
constant ($H_0$) is assumed. \cite{mac05} 
 defined \sigh\  by the   standard deviation (SDV) of the residual LOS
 velocities. This is equivalent to measuring the  RMS of the residual
 velocities from the local Hubble flow, namely using the locally
 determined Hubble constant from the data itself ($H_\mathrm{loc}$).
 Yet, a simpler approach is to associate \sigh\ with the RMS of the
 residual from the global Hubble flow. This is the approach adopted
 here, namely \sigh\ is calculated by: 
\begin{equation}
\label{eq:sigma-H}
\sigma{^2_H}= {1\over N} \sum_{i=1}^N{[v_i-H_0r_i]^2}.
\end{equation}
and the sum extends over the $N$  galaxies within 
 the appropriate distance cut of the catalog. Hereafter these methods are referred to as the APEX, SDV and RMS.

\begin{figure}
\bc
 \resizebox{8.5cm}{!}{\includegraphics{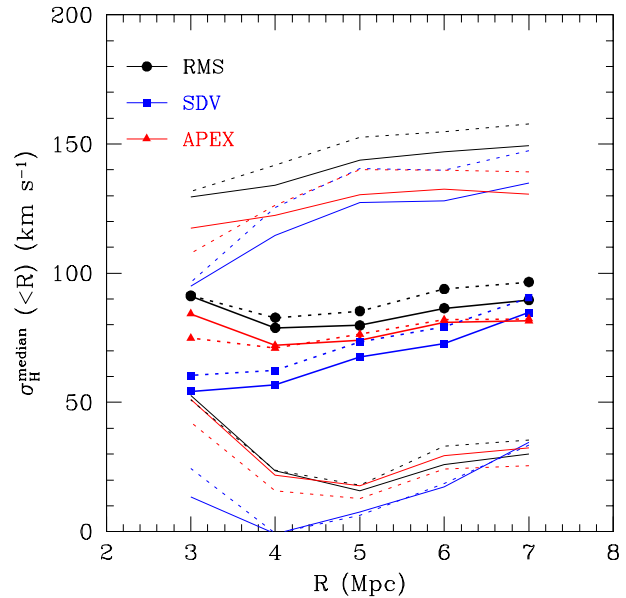}}
\caption{The median of \sigh\ (lines with points) and its  standard deviation 
   (thin lines)  for the \LCDMhr\ simulation taking all the
  DM halos (solid lines) and only those with a circular velocity higher
  than $35 \kms$ (dotted lines) in the distance range of $(1 -  R) \mpc$
  computed using the APEX (triangles), SDV (squares) and RMS (circles) methods. 
}
\label{fig1}
\ec
\end{figure}

One can argue for the merits and disadvantages of the different
methods and  they all can be used as long as observations and
simulations are analyzed consistently. Still, we argue here that the
use of the dispersion around the global Hubble flow is the preferred
way. The motivation for measuring \sigh\ stems from the information it
provides on the deviations from a pure Hubble flow. By using the
locally determined Hubble constant the 'breathing' (i.e. isotropic)
mode  of the flow is absorbed in $H_\mathrm{loc}$, and so the resulting
\sigh\ underestimates the actual deviation from a pure Hubble flow.
 In the case of the apex motion the velocity of the LG, that hosts the
 observer, is treated as if it is an external effect that is not
 related to the perturbed flow one is trying to study. It is solved in
 a manner which does  not depend on the nature of the perturbation
 field. The inclusion of the apex motion of the observers provides a
 better measure of the deviation from a pure Hubble flow. Yet, it is
 shown here that the differences between the different measures are
 small. 
The different measures of \sigh\ have been applied to an ensemble of LG-like objects drawn from one of the high resolution simulations studied here (\LCDMhr, see \S\ref{sec:sim} for a detailed description). Figure \ref{fig1} presents the  results obtained for the different estimators of \sigh. The main result is that the differences between the different methods are much smaller than the scatter around the median of calculated values of \sigh.

Table 2 of   \cite{tik08} provides the RMS and APEX estimates  of
\sigh\ for  the CNG galaxies for various radial cuts. These estimates
are corrected for  distance errors. The representative number used here is the RMS estimation for all galaxies within $(1 \  - \ 7)\mpc$ of $\sigma_H=90.4 \kms$.

\section{Simulations}
\label{sec:sim}

A suit of constrained and unconstrained \LCDM\ (flat, $\Lambda$ and CDM),
OCDM (open, CDM only) and SCDM (flat, CDM only) low and high resolution
simulations have been performed. Table \ref{tab:models} lists the names
and the parameters of the simulations. The five low resolution
simulations ($N=256^3$) are the same ones described in \cite{mv07} and
\cite{hf08}. 
A computational box of $L=64\hmpc$ (where $h$ is Hubble's constant in units of $100 \kms \mpc^{-1}$) and the WMAP1 cosmological parameters are assumed for the \LCDM\ simulations \citep{wmap1} .
The OCDM simulations are identical to the \LCDM\ ones but without the
$\Lambda$ term, hence they correspond to an open universe.  
The SCDM is used here just as a extreme  model  for which the Hubble
flow is expected to be much hotter than all other models,  despite the
fact that  is a cosmological model which stands 
in marked disagreement with a wide range of cosmological observations.
The five low resolution,  constrained and unconstrained, simulations are all based on the same  realization of the Gaussian random field. 
The \HRLCDM\ and \LCDM160 are high resolution ($N=1024^3$) simulations
in  computational boxes of $L=64$ and $160 \hmpc$ respectively. Both 
are performed with the WMAP3 cosmological parameters \citep{wmap3} .

Five of the simulations are constrained ones, namely the initial conditions of these are set by constrained realizations of Gaussian fields. Observational data of the nearby universe is used as constraints on the initial conditions and thereby the resulting simulations reproduce the observed LSS.
The basic algorithm of constraining Gaussian random fields is the one developed by Hoffman \& Ribak (1991). The implementation of the algorithm to observational data and a description of the construction of constrained simulations was described at length in \cite{kra02} and \cite{kly03}. 
A brief description of the constraining data is given here. Two different observational data is used to set up the initial conditions. The first   is made of radial velocities of
galaxies drawn from the MARK III \citep{mark}, SBF \citep{sbf} and
the \cite{kar04} catalogs. Peculiar velocities are less affected by
non-linear effects and are used as constraints as if they were linear
quantities \citep{hof99}.  The other constraints are
obtained from the catalog of nearby X-ray selected clusters of galaxies
\citep{rei02}. Given the virial parameters of a cluster and assuming
the spherical top-hat model one can derive the linear overdensity of
the cluster. The estimated linear overdensity is imposed on the mass
scale of the cluster as a constraint. 
For the CDM cosmogonies the data used here constrains the simulations
on scales larger than $\approx 5 \hmpc$ \citep{kly03}. It follows that the
main features that characterize the local universe, such as the Local
Supercluster, Virgo cluster,  Coma  cluster and  Great attractor (in the large box), are all reproduced by the simulations. The small scale structure is hardly affected by the constraints and is essentially random.

We have used the parallel TREEPM N-body code  GADGET2 \citep{gadget2} 
to run these simulations. An  uniform mesh of $512^3$  grid points was
used to compute the long-range gravitational force by means of the
Particle-Mesh algorithm. A constant comoving Plummer equivalent 
 gravitational smoothing scale of $20 \hkpc$  was set  at high
 redshift and we changed it to  $5 \hkpc$  kpc physical scale 
since z=3 till z=0.  The simulations were started at z=50 in all cases.
We employed a variety of  parallel computer architectures
(SGI-ALTIX, IBM-SP4, Opteron-clusters) during
the course of this work. Using  16 processors simultaneously, we
completed  one run in  about 2 cpu days. 

We analyzed also two simulations with higher mass resolution ($1024^3$ particles). 
The  simulation with box size $64 \hmpc$  has been also evolved using
GADGET2. A grid of $1024^3$ was used in this case for the computation
of the PM forces. Gravitational smoothing was set to $1.6  \hkpc $  at
high redshift and $0.8 \hkpc$ at lower redshift. The initial conditions
were set up  at $z=100$ for this run.  The  simulation of box size $160 \hmpc$
has been 
calculated by 
 the MPI version of the Adaptive Refinement Tree
(ART) code described in \cite{ART}. Initial conditions were set up at
$z=30$ and the maximum number of refinement levels were set to 9,
which translates into a maximum  spatial resolution of $ 1.2  \hkpc$.
 These two high resolution
simulations were also  analyzed by \cite{tik08}.

We conclude this section  with a general remark. The parameters of the
simulations are defined by using the $h^{-1}$ scaling. Yet, the
comparison of the results with observations, and in particular the
distance cuts, is made after an $H_0=73 \kms \mpc^{-1}$ is assumed and
distances are expressed in units of ``real'' $\mpc$.

\begin{table*}
  \begin{center}
    \begin{tabular}{ccccccccc}
      \hline
      MODEL & CONSTRAINED & BOX [$\hmpc$]  & $\Omega_m$ & $\Omega_\Lambda$ & h & $\sigma_8$  & $N$  & $m_{DM} [\hmsun] $ \\ 
      \hline
      \LCDM\       & yes  &   64 & 0.30   & 0.70   & 0.70    &  0.90    & 256$^3$   &  $1.3 \times 10^9$  \\
      \LCDMu      &  no       &   64 & 0.30   & 0.70   & 0.70    &  0.90    & 256$^3$   &   $1.3 \times 10^9$  \\
      \HRLCDM\    & yes  &  64 &  0.24  &  0.76  &  0.73    &  0.75   & 1024$^3$ &   $1.6 \times 10^7$   \\
      \LCDM160 & yes  &160  & 0.24  &  0.76  &  0.73    &  0.75   & 1024$^3$ &    $2.5 \times 10^8$   \\
      OCDM        & yes  & 64   & 0.30   & 0        & 0.70     &  0.90    & 256$^3$   &   $1.3 \times 10^9$  \\
      OCDMu     &   no      &  64   & 0.30   & 0        & 0.70     &  0.90    & 256$^3$   &   $1.3 \times 10^9$   \\
      SCDM        & yes  &64     & 1.0      & 0        & 0.50     & 0.7       & 256$^3$   &    $4.4 \times 10^9$  \\
      \hline
    \end{tabular}
 \caption{
 Description of the set-up and cosmological parameters used for the
 different simulations: constrained or  random, computational box size,
 matter density ($\Omega_m$), cosmological constant ($\Omega_\Lambda$), Hubble's
 constant ($h$), initial
 power spectrum normalization ($\sigma_8$), number of particles ($N$) and
 mass of the DM particle ($m_{DM}$) 
 }
   \label{tab:models}
  \end{center}
\end{table*}

\section{Selection of LG  Candidates}
\label{sec:sel}

\begin{table*} 
  \begin{center}
    \begin{tabular}{cccccccc}
      \hline
      Criterion & \LCDM\ & \LCDMu & \HRLCDM\ & \LCDM160 & OCDM & OCDMu & SCDM \\ 
      \hline
      \pairs\    & 13 & 20 & 12 & 131 & 11 & 15 & 24 \\
      \singles\  & 52 & 65 & 43 & 478 & 38 & 60 & 78 \\
      \hline
    \end{tabular}
    \caption{Number of LG candidates found in each simulation following
      the \pairs\  and \singles\  criteria.}
    \label{tabcandi}
  \end{center}
\end{table*}

DM  halos were found in simulations using two object finding
methods: The Bound Density Maxima (BDM) algorithm
\citep{bdm}  is   based on finding local center of mass  in spheres of variable radius
starting from randomly selected particles in the simulation. 
 The AMIGA Halo Finder 
 \citep{amiga},  on the contrary, finds local
density  maxima  from an adaptive mesh hierachy. 
In both cases, an iterative procedure to find local centre of mass from
density maxima is used. Particles  that are  gravitationally unbound to
the halo potential are also removed.
Only halos with a mass higher than $2.6 \times 10^{10}$ (\LCDM, \LCDM
u, OCDM and OCDMu), $8.7 \times 10^{10}$ (SCDM), $5.0 \times 10^{8}$
(\HRLCDM) and $4.0 \times 10^{9} M_\odot $ (\LCDM160) are considered.
For the work
reported here we have used the halo catalogues obtained  by the AMIGA code, 
except for the \LCDM160 simulation where we have used the BDM halo
catalogue. In any case, we have checked that  the results are
independent of the halo finder. 

One of the main aims of this paper  is to perform a detailed comparison
with  \cite{mac05} results, so as to be able to challenge their claim on the role of the DE in the local dynamics. In order to meet this goal the selection criteria of  LG-like objects of  \citeauthor{mac05} have been followed to the letter. 
Going beyond the comparison with that paper the selection rules are reexamined and some of which are revised.
Here, these criteria are expressed in physical units without the $h^{-1}$ scaling.
The LG objects obey the following selection:

\begin{enumerate}
 \item The groups contain two MW and M31 like DM halos with maximum circular velocity in the range of $ 125 \leq V_c \leq 270 \,\rm km/s$.
 \item The two major DM halos are separated by no more than  $1\mpc$.
 \item The relative radial velocity of the two main halos is  negative.
 \item There are no objects with maximum circular velocity higher than MW and M31 candidates
 within a distance of $3\mpc$.
 \item The group resides within a distance of $7$ to $17\mpc$ from
 one and only one 
Virgo like halo of $ 500 \leq V_c \leq 1500 \,\rm km/s$.
No Virgo like halos can appear within a distance of $7\mpc$.
\end{enumerate}

 The selected  LG-like objects obeying  all  the above 
 rules are defined as \pairs, for the fact that they are dominated by   two MW and M31 like DM halos.   
 
A careful examination  of the dynamics of LG-like objects suggests the dynamics of the Hubble flow might not depend strongly on whether they are dominated by two, almost equal, massive objects or by rather one massive objects (see \citeauthor{mac05} \citeyear{mac05} and \citeauthor{tik08} \citeyear{tik08}). To test this idea we construct the \singles\  ensemble of LG-like objects which obey all the criteria presented above apart from the first one. Namely, for  the \singles\  objects we look for individual halos, 
whose mass are similar to MW and M31 together, namely $ 10^{12} \leq M_{vir} \leq 2 \times 10^{12} \hmsun$. 
The LV around such objects is studied.

In Table \ref{tabcandi}, the number of LG candidates found following both criteria is shown.
Higher number of LG-like objects are obtained using \singles\  criterion becuse it is less restrictive than
the \pairs\  one. 
One should note that there is not a one-to-one correspondence of mock
LGs in the \LCDM\ and OCDM simulations, even if they have the same
computational box and random realization of the initial
conditions. This stems for the fact that the LG constitutes a
quasi-linear object, far from being in virial equilibrium. LG-like
objects are delicately defined to match the observed LG and the small
dynamical differences  introduced by the $\Lambda$ term are likely to prevent
a full correspondence between objects in the different models.

Some of the simulations used here are constrained ones, namely their large scale structure is constrained to reproduce the observed cosmological neighborhood. 
No attempt is made  here to select only LG-like objects that reside in the 'correct' position within the cosmic web that constitutes the local neighborhood, as this would result in a very poor statistics. Rather,   objects are defined as LG-like, \pairs\ or \singles,
regardless of their location in the computational box, and the flow field around these objects is studied.

\section{Analysis of Hubble Flow within the Local Volume}
\label{sec:result}

\begin{figure*}
\bc
\resizebox{15cm}{!}{\includegraphics*{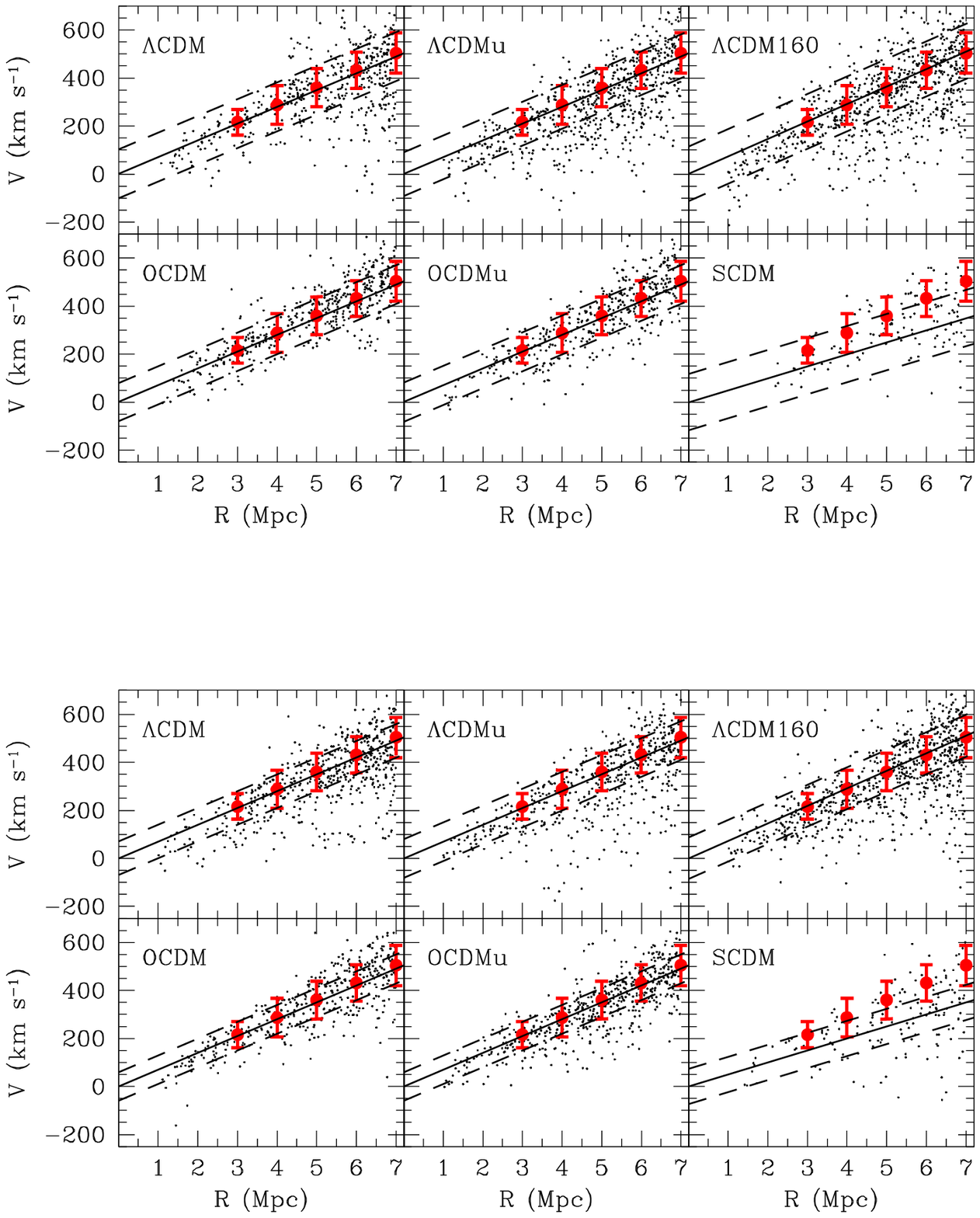}}
\resizebox{15cm}{!}{\includegraphics*{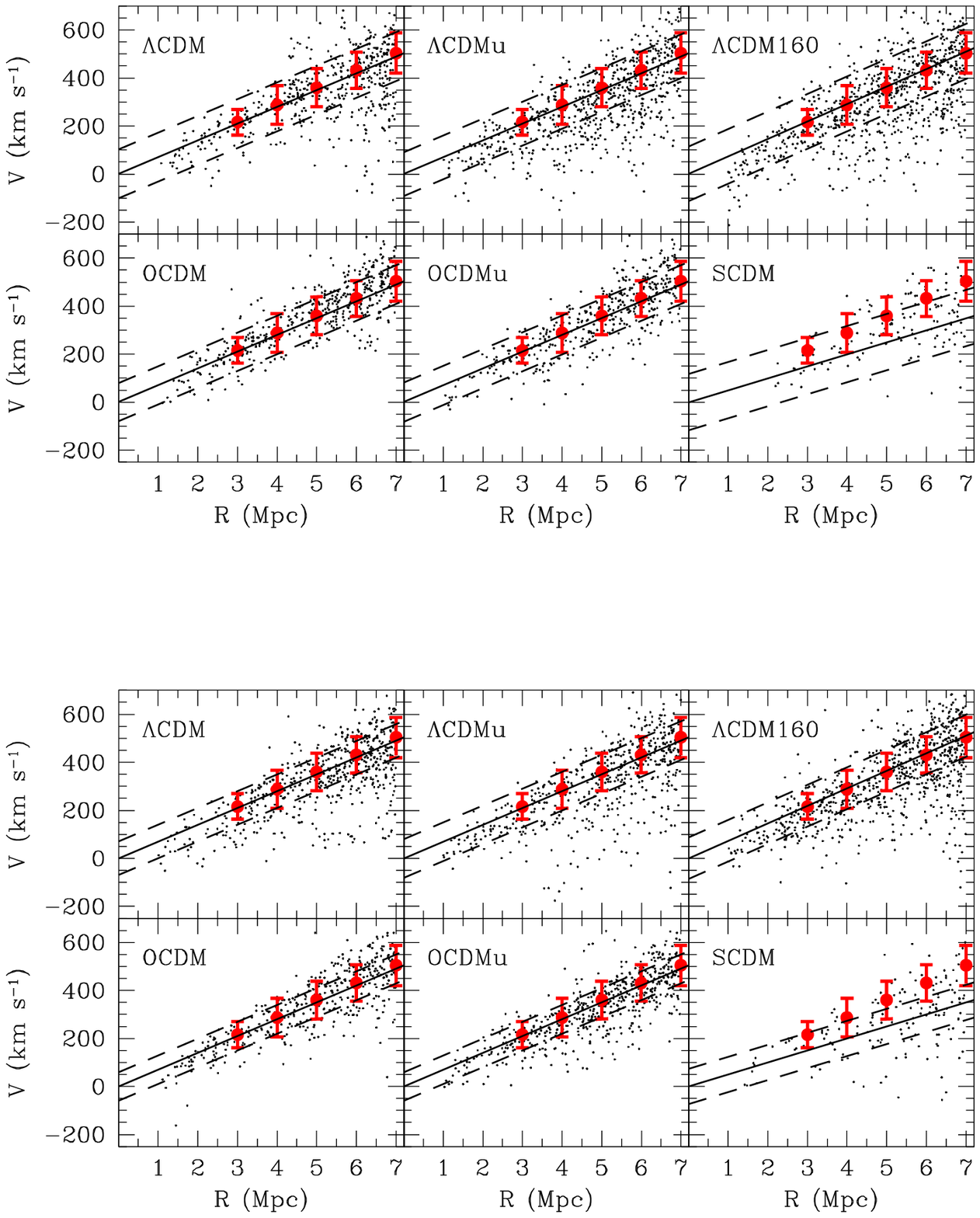}}
\caption {Combined Hubble diagrams of 10 randomly chosen LG-like
  objects  in the different simulations. The solid lines represent the
  $H_0\  r$ Hubble flow and the dashed ones show the median of \sigh\
  (see Table \ref{tabsig}). The individual data points (in red)
  correspond to the mean radial  velocity and the \sigh\ dispersion in
  the $(1 -  R) \mpc$ 
  distance cut of the Tikhonov \& Klypin (2008) data. The
  upper panel presents the LG-like objects selected using the \pairs\
  criterion  and the lower ones  correspond to those LG's selected with the \singles.}
\label{fig2}
\ec
\end{figure*}

\begin{table*} 
  \begin{center}
    \begin{tabular}{cccccccc}
      \hline
      Criterion & \LCDM\ & \LCDMu & \HRLCDM\ & \LCDM160 & OCDM & OCDMu & SCDM \\ 
      \hline
      \pairs\ & 102 $\pm$ 40 & 91 $\pm$ 47 & 90 $\pm$ 60 & 114 $\pm$ 58 & 80 $\pm$ 25 & 81 $\pm$ 30 & 117 $\pm$ 33 \\
      \singles\ & 70 $\pm$ 41 & 80 $\pm$ 44 & 69 $\pm$ 42 & 87 $\pm$ 46 & 60 $\pm$ 22 & 60 $\pm$ 30 & 74 $\pm$ 38 \\
      \hline
    \end{tabular}
    \caption{The median  of  \sigh\ (in $\kms$ units) and the standard
      deviation  from the median for all the LG candidates in each
      simulation  selected  with  \pairs\  and \singles\  criteria (see Fig. \ref{fig2}).}
    \label{tabsig}
  \end{center}
\end{table*}

\begin{figure*}
\bc
\resizebox{15cm}{!}{\includegraphics{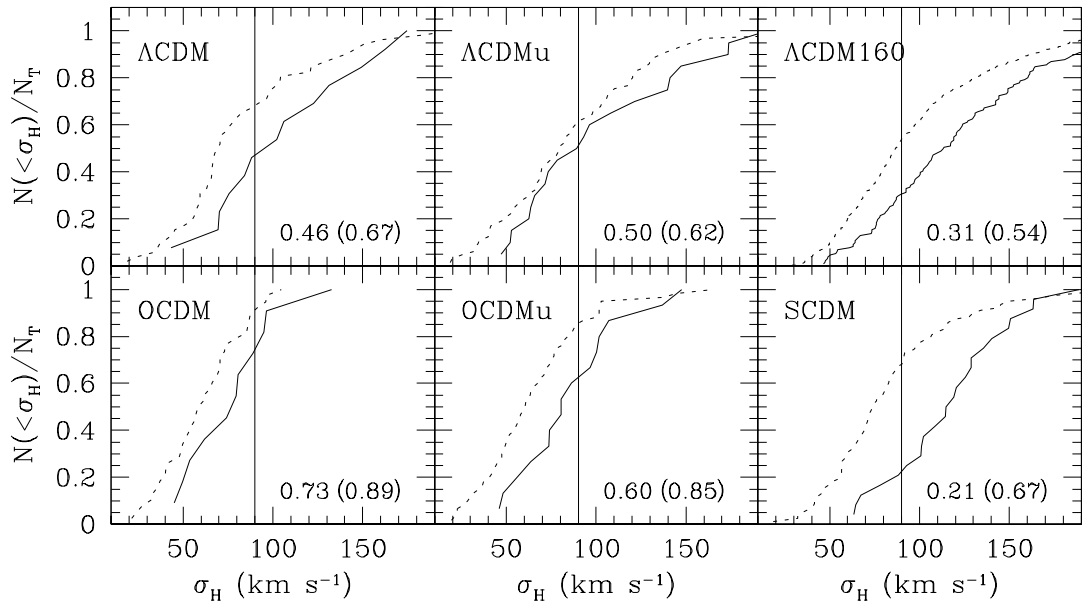}}
\caption { 
Cumulative number of LG candidates, from \pairs\  (solid lines) and
\singles\  (dashed lines) criteria, with a \sigh\ lower than a given
value, for the different simulations. The vertical line shows the value
of the observational $\sigma_H=90.4\kms$. The numbers on each frame are the
fraction of LG-like objects with \sigh\ below this observational value
for the \pairs\ and \singles\ (in parentheses). 
}
\label{fig3}
\ec
\end{figure*}

\begin{figure*}
\bc
\resizebox{15cm}{!}{\includegraphics*{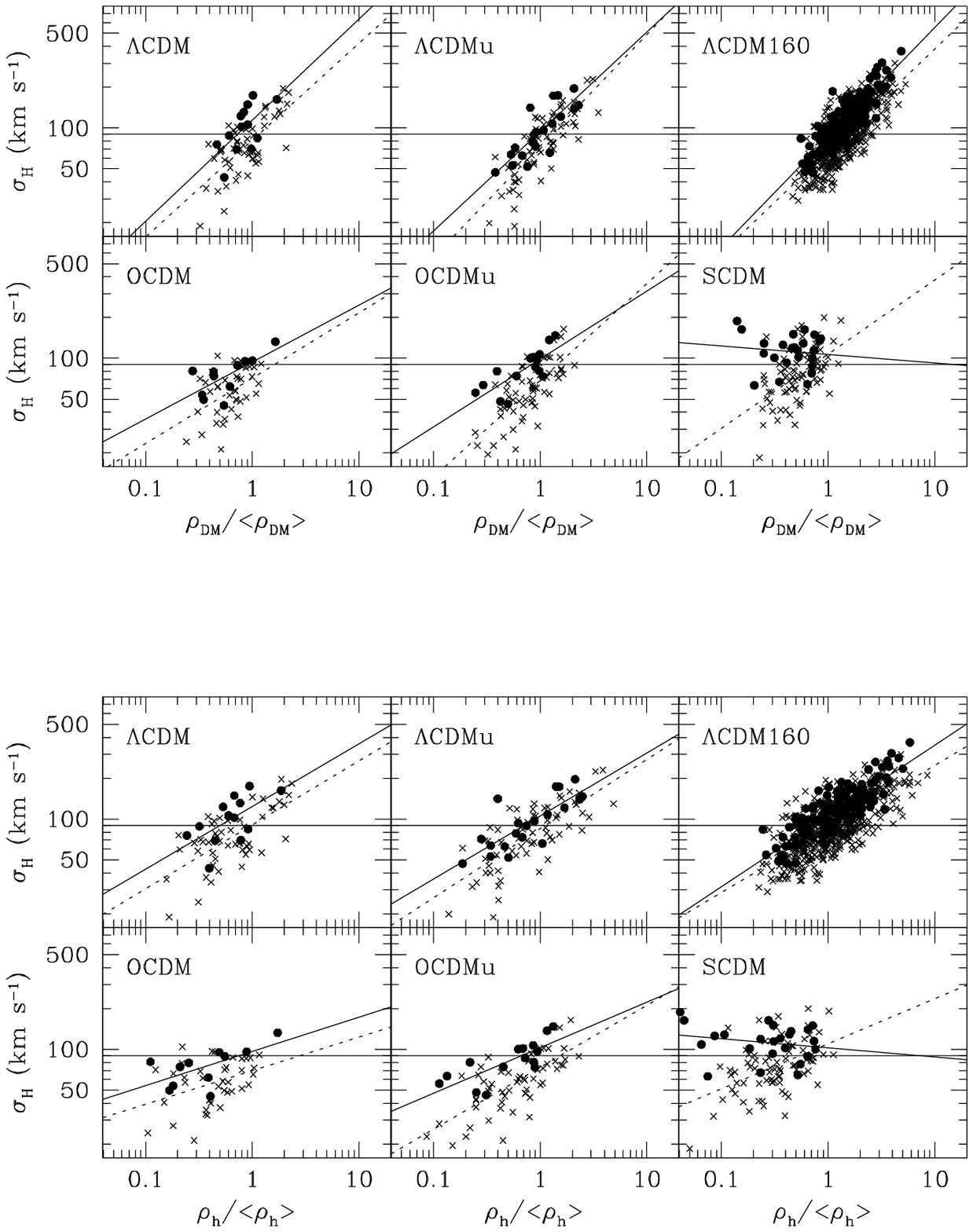}}
\resizebox{15cm}{!}{\includegraphics*{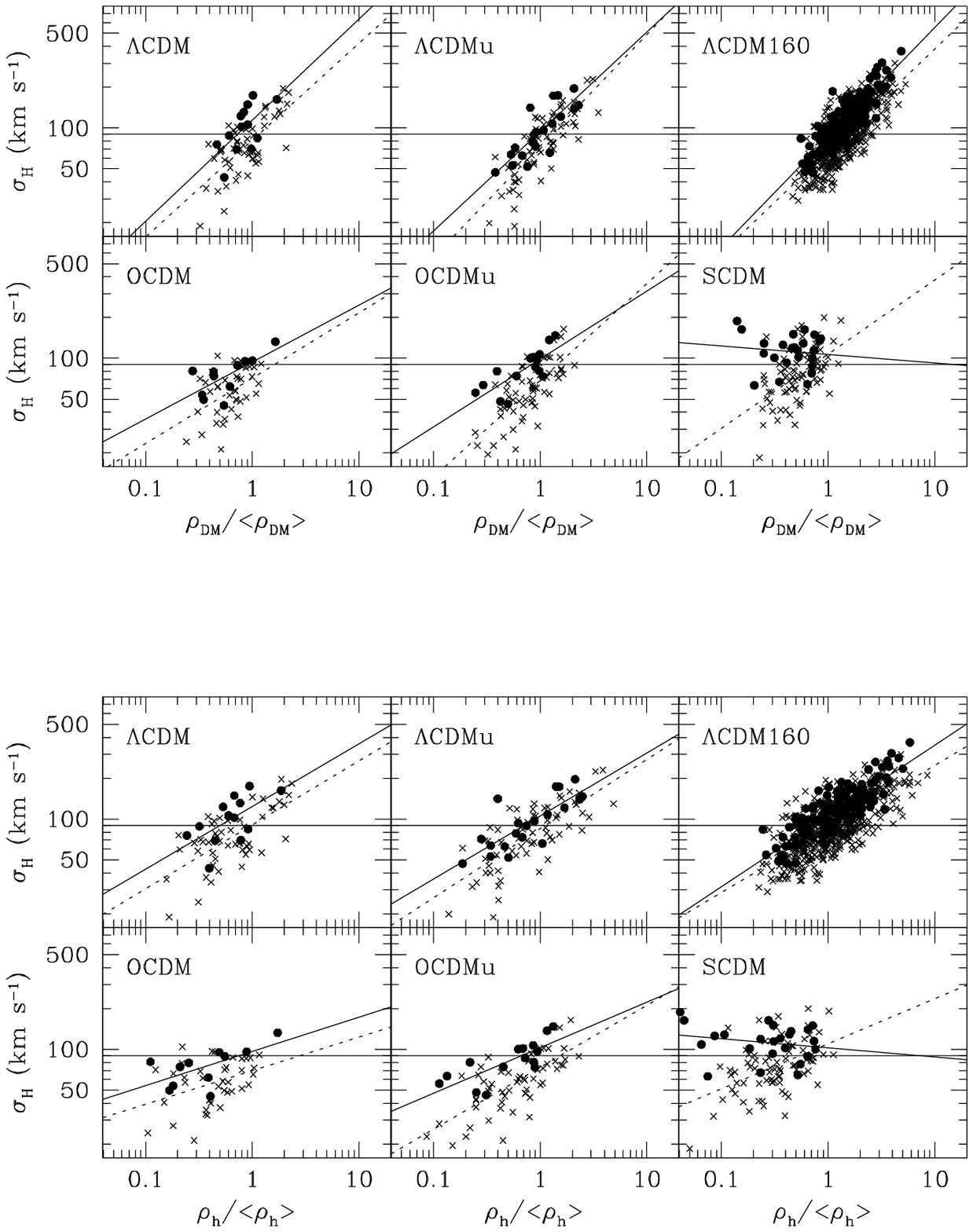}}
\caption {
A scatter plot of the relation between \sigh\   and the mean matter density
(normalized by the mean cosmological density) within the LV ($R=7\mpc$)
for \pairs\ (solid points) and \singles\ (crosses).  Power law fits to the
\sigh\ - density scatter are shown for the \pairs\ (solid line) and
\singles (dashed line).   The horizontal line indicates  the
observational \sigh. 
In the upper panel the matter density is calculated from all  DM
particles inside the sphere. In the lower panel, 
 the density is calculated from DM  particles belonging to 
halos only.
}
\label{fig4}
\ec
\end{figure*}

The main purpose of the present section is to study the dispersion of the peculiar velocities around LG-like objects selected in the manner of  \cite{mac05}. 
The flow around each object is studied using the RMS estimator of \sigh, so as to allow a comparison with the analysis of the CNG data \citep{tik06}.
The analysis consists of three parts: 
\begin{itemize}
\item[(a)] The examination of the Hubble diagram of the flow around
  LG-like objects.
\item[(b)] The statistical distribution of $\sigma_H$ of the various models. 
\item[(c)] The dependence of $\sigma_H$ on the mean density of the LV around
  each object. 
\end{itemize}
In all cases we take a  spherical shell of $(1 \  - \ 7) \mpc$ as the
LV of each object and, unless  otherwise stated,  $\sigma_H$ corresponds to all objects within
the LV. Like with the actual data, the inclusion or omission of  halos
within  $1.0 \mpc$ distance leaves  the value of $\sigma_H$ virtually unchanged.

The six panels of Fig. \ref{fig2} present the combined Hubble diagram
of 10 randomly chosen LG-like objects of the different models. The
solid line corresponds to the unperturbed Hubble flow and the upper and
lower dashed lines corresponds to $H_0 r \pm \sigma_H$, where $\sigma_H$ is the
median value of $\sigma_H$ for each simulation.  The individual points with
the error bars represent the value of $\sigma_H(R)$ from \cite{tik08}
observational data. 
The results from  \HRLCDM\ simulation are  not shown in Figures
\ref{fig2}, \ref{fig3} and \ref{fig4} since  they are very similar to
 the other \LCDM\ simulations.

 Table \ref{tabsig} presents the mean $\sigma_H$ of the full LV  taken over the \pairs\  and the \singles\  in each of the simulations.
The fractional cumulative distribution, $\eta(\sigma_H)= N( < \sigma_H)/N_T$, is
presented in  Fig. \ref{fig3}, where $ N( < \sigma_H)$ is the number of
LG-like objects colder than \sigh\ and $N_T$ is the total number of LG
objects. The cumulative distribution is shown  for all models and
for both the \pairs\  and the \singles. In each frame the vertical line
indicates the observational value of $\sigma_H= 90.4 \kms$ and the numbers
give the fraction of LG-like objects with \sigh\ smaller than the
observed value of the \pairs\ and the \singles\ (in  parentheses).

 Fig. \ref{fig3}  shows that there is a large scatter in the
 distribution of $\sigma_H$ and that for all the models considered here
 there is a non-negligible   probability to find LG-like objects with
 $\sigma_H$ equal or smaller the one corresponding to the actual
 LG. However, there is another dynamical variable that controls the
 dynamics around these objects and that is the mean density within the
 LV. This was  shown by \cite{mac05} and it is clearly reproduced by
 the analysis of the models considered here.  Fig. \ref{fig4} shows the
 scatter plot of $\sigma_H$ {\it vs.} the mean matter density within the LV
 normalized by the mean cosmological density. Both \pairs\  and
 \singles\  LG-like objects are studies and the results are virtually
 the same in the range of $\rho_\mathrm{DM}/\bar{\rho}_\mathrm{DM} \leq 2$,
 where $\rho_\mathrm{DM}$ and $\bar{\rho}_\mathrm{DM}$ are the mean DM
 density within the LV and the cosmological mean DM density.  
In the simulations the DM density is very easily measured but this is
not a directly observable quantity. Observationally the galaxy
distribution within the LV is used to infer a local density. Short of
having numerical simulations with full galaxy formation treatment the
density of DM halos is used here as a proxy to the galaxy
distribution. 
The lower panel of  Fig. \ref{fig4} presents the
dependence of $\sigma_H$ on the density of mass which belongs to halos
within the LV normalized by the mean mass density of all the simulated
volume. In both panels of Fig. \ref{fig4}  the horizontal line denotes the 
observational estimate of \sigh.

\section{Relaxing the Local Group Criteria}
\label{sec:relax}

\begin{table*} 
  \begin{center}
    \begin{tabular}{ccccc}
      \hline
       & Distance between members (Mpc) & No neighbours (Mpc) & Virgo clusters & \sigh\ ($\kms$)  \\ 
      \hline
      a &1.0 & 3 & only one & 102 $\pm$ 40  \\
      b & singles & 3 & only one & 70 $\pm$ 41  \\
      c & 1.5 & 3 & only one & 85 $\pm$ 47  \\
      d & 1.0 & no constrain & only one & 189 $\pm$ 100  \\
      e & 1.0 & 5 & only one & 104 $\pm$ 14 \\
      f & 1.0 & 3 & no constrain & 113 $\pm$ 112  \\
      g & 1.0 & 3 & one or more & 150 $\pm$ 74 \\
      \hline
    \end{tabular}
    \caption{The median of  \sigh\ and the standard deviation  from the
      median  for the candidates in the \LCDM\ simulation for different
      selection criteria (see text).}
    \label{tabrelax}
  \end{center}
\end{table*}

\begin{figure}
\bc
\resizebox{8.5cm}{!}{\includegraphics{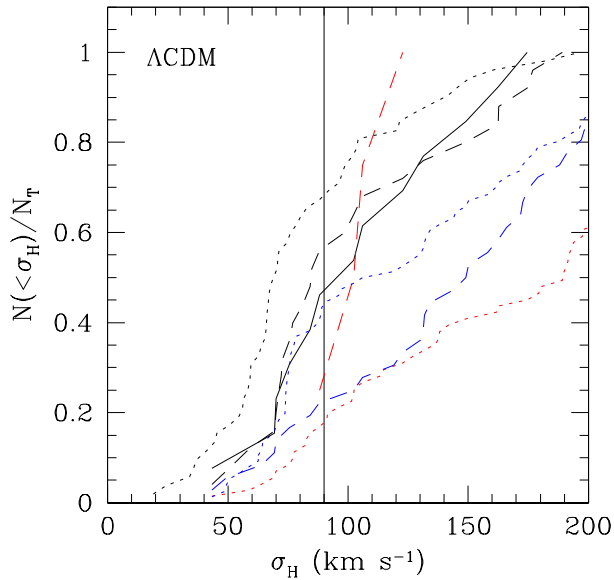}}
\caption { Fractional cumulative distribution  of   \sigh\  for  
the  LG candidates in  the \LCDM\ simulation found with the different
criteria shown is Table \ref{tabrelax}.  The black lines
  correspond to \textit{a} (solid), \textit{b} (dotted) and \textit{c}
  (dashed), the red ones to \textit{d} (dotted) and \textit{e}
  (dashed), and the blue ones to \textit{f} (dotted) and \textit{g}
  (dashed) criteria. The vertical line shows the value of the
  observational \sigh.} 
\label{fig5}
\ec
\end{figure}

The LG appears to be a very typical small group of galaxies, whose main  dynamical characteristics   are summarized in \S\ref{sec:sel}. These are formulated in terms of the five criteria for the selection of LG-like objects. 
These dynamical properties are common in the Universe, yet the
particular dynamical configuration of the LG determines  \sigh. It is
interesting to see how these properties affect the flow field around
the LG and for that purpose we have relaxed some of the criteria of
\S\ref{sec:sel}, reselected  ensembles of mock LGs  and studied the
cumulative histogram of \sigh\ for the various selections 
(in the \LCDM\ simulation).
This is shown in Table \ref{tabrelax} and Fig. \ref{fig5}, where the
median and the cummulative fraction of LG-like objects of the
constrained \LCDM\ simulation,  respectively, is shown for the
following cases: 
\begin{itemize}
\item[(a)] The \pairs, which are taken as a benchmark.
\item[(b)] The case of \singles.
\item[(c)] The \pairs\  criteria, but assuming a distance between LG members lower than $1.5\mpc$.
\item[(d)] No constraint is imposed on the lack of nearby galactic size halos.
\item[(e)] There must not be  objects similar to LG members within a distance of $5\mpc$.
\item[(f)] 
No constrained is imposed on the existence of a Virgo-like cluster. 
\item[(g)] There must be one or more Virgo-like clusters. 
\end{itemize}

The \pairs\ selection is taken here as a benchmark and it is represented here for the sake of completeness.
The largest departure from the benchmark \pairs\ is obtained by removing the constraint on nearby LG-like objects (case d).
Table \ref{tabrelax} and Fig. \ref{fig5} clearly show the role of
clusters in heating the flow.   A nearby Virgo-like
cluster exerts a tidal field in the vicinity of the LG-like object,
resulting in a shear flow which contributes to the anisotropic
component of the velocity field and thereby heating the flow. The
benchmark case (a)  has only one Virgo-like cluster while in case (g)
there can
be  one or more   clusters. As expected, the flow for
objects found in case (g) is 
significantly 
hotter than for those found following case (a). 
The removal of the Virgo constraint increases somewhat 
the median but it more than doubles the standard deviation (case f). 
In such a case mock LGs 
can either have rich clusters within their LV, hence with very high \sigh, or 
can reside in low density regions far away from clusters and
consequently have very low \sigh. 
Similarly, allowing for nearby MW size halos
heats the flow and 
leads to many  hot LG-like systems
(case d). 
 The analysis shows that the most
significant selection criterium  is that of the nearby halos (d). By
omitting this  criterion only roughly 15\%  of the LG-like objects have
\sigh\  smaller than the observed value.

\section{Mass Resolution}
\label{sec:res}

The numerical simulations used here span a mass resolution that ranges
over more than two orders of magnitude. The majority of the simulations
are done on a very coarse grid of $256^3$ and have  low mass
resolution. In fact, the LG-like objects of the low resolution \LCDM\
and OCDM are made of about a thousand DM particles and the minimal mass
of  DM halos is $\approx 3\times 10^{10} \hmsun$. The faint nearby galaxies  are
presumably much less massive than this lower limit and the question
arises as to what extent the present results are affected by the lack
of resolution. 

\begin{figure}
\bc
\resizebox{8.5cm}{!}{\includegraphics{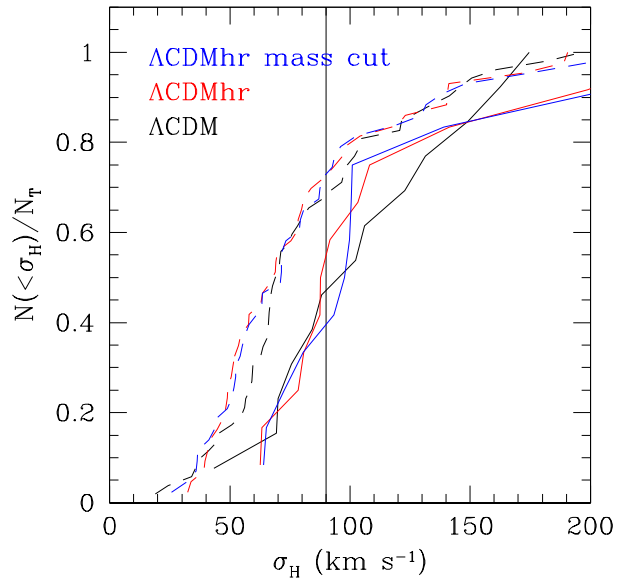}}
\caption { The fractional cumulative distribution  ($\eta(\sigma_H)$),  of LG
  candidates for  simulations with  different mass resolutions:  
The black curves correspond to the \LCDM\  simulation 
and the red ones to the \HRLCDM.
The blue curves correspond to  \HRLCDM, but using only halos within the
LG's  in the same mass limit than those   in the \LCDM\ simulation.  
The solid lines  correspond to the LG-like objects selecte by  the \pairs\ and the dashed ones
to those selected by the \singles\ criteria. 
The vertical line show the observational $\sigma_H=90.4\kms$. }
\label{fig6}
\ec
\end{figure}

To address the issue of mass resolution we have analyzed the \HRLCDM\
simulation and compared it with the \LCDM\ one. This is a high
resolution ($N=1024^3$) \LCDM\ (WMAP3) constrained simulation in the
same  64 $h^{-1} $ Mc volume. This simulation differs from the low resolution \LCDM\ one
in terms of resolution, cosmological parameters and hence the power
spectrum. It constitutes a different realization of the random
field used to set the constrained initial conditions. An ensemble of
LG-like objects has been constructed and \sigh\   has been calculated
for all the DM halos in the corresponding LV (Figure \ref{fig6}, red
curves) and for DM halos above the mass limit of the \LCDM\ simulation
($\approx  3\times 10^{10} \hmsun$; blue curves).
These are compared with the cumulative distribution of the low
resolution \LCDM\ simulation. 
 Figure  \ref{fig6} shows very clearly  that, within the range studied
here, the distribution of \sigh\ does not depend on the mass
resolution.

\section{The Local Volume and Blueshifted Galaxies}
\label{sec:blue}

A striking manifestation of the coldness of the local flow   is the
absence of blueshifted galaxies in the local neighborhood. In
particular the CNG catalog shows only one blueshifted  galaxy in the
distance range of $(1 \  - \ 7 )\mpc$, and it is flagged as a probable
member  of the Virgo, i.e. its distance might be erroneous.  
The absence of blueshifted galaxies provides only a qualitative measure
for the coldness of flow, yet it constitutes a very selective
criterion for finding  LG-like objects \citep{sch94} .
The implications of such a  selection on the coldness of the flow around the selected groups are studied here.

\begin{figure*}
\bc
\resizebox{12cm}{!}{\includegraphics*{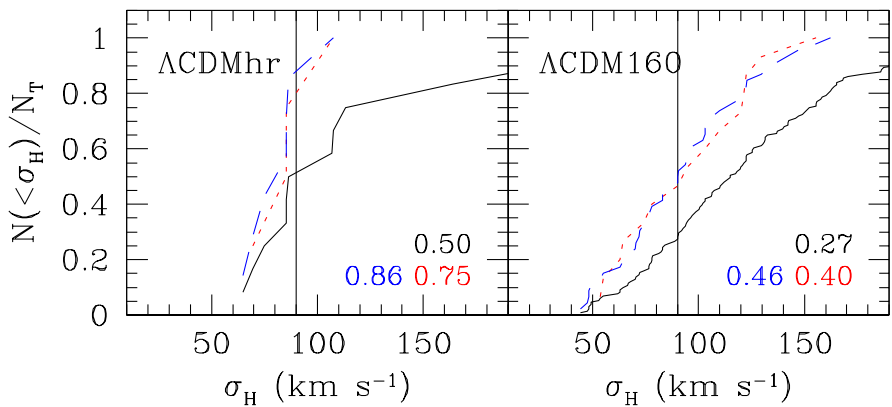}}
\caption {
The Fractional cumulative distribution  of LG candidates ($\eta(\sigma_H)$)
for the two high resolution \LCDMhr\ 
(left) and \LCDM160 (right) simulations.   
the solid line corresponds to the  \pairs\ criterion, the other two  lines
represents groups with no blueshifted DM halos with $V_{circ} > 35 \kms$ within
 5 -- 7 $ \mpc$ (red-dotted) and 3 -- 7  $\mpc$ (blue-dashed)
shells. The vertical line shows the observational $\sigma_H=90.4 \kms$. 
}
\label{fig7}
\ec
\end{figure*}

\begin{figure*}
\bc
\resizebox{12cm}{!}{\includegraphics*{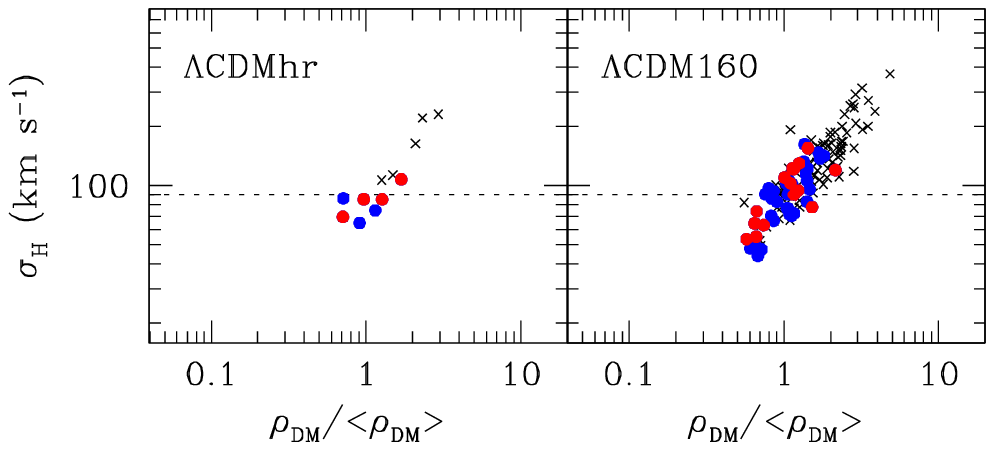}}
\caption {
Scatter plots of \sigh\ vs. the mean matter density, normalized by the
mean cosmological density,   within the LV's found in the high
resolution simulations,  \LCDMhr\ (left) and \LCDM160
(right).  
All symbols correspond to LV's around LG-like objects found with the
\pairs\ criterion in both simulations.  The filled circles  represent
LV's  with  no blueshifted DM halos with $V_{circ} >
35 \kms$ within the 5 -- 7  Mpc shell (blue)  and   within 3 -- 7 Mpc
shell (red). The crosses are for the rest of the LV's.  
}
\label{fig8}
\ec
\end{figure*}

A word of cautious is due before applying the constraint of no
blueshifted 'galaxies' to the LG-like objects. Here, the local velocity
field is traced  by DM halos. It follows that the number of blueshifted
objects might depend on the  mass resolution of  the simulations. 
We follow here 
\cite{tik08} and assume  that in the \LCDM\ model DM halos with
$V_{circ} > 35 \kms$ serve as proxy for observed luminous galaxies in
the LV.   
Such  halos are well resolved in the  two high resolution simulations
to which  the no blueshifted 'galaxies'  selection is applied.  Table
\ref{table:blue} lists the results of \sigh\ analysis applied to the
\LCDMhr\ and \LCDM160 simulations. The analysis is applied to DM halos
with $V_{circ} > 35 \kms$. The analysis is applied to all the LG-like
objects and then to groups that contain no blueshifted galaxies in the
shell corresponding to $(x \ - \ 7)\mpc$, where $x= 2, 3 $ and $5
\mpc$. The Table lists the number of LG-like objects and the median and
standard deviation of \sigh\ of the selected objects. Figure \ref{fig7}
shows the cumulative histogram of \sigh\ for the \LCDMhr\ and \LCDM160\
simulations. The figure presents the benchmark case of all pairs and
the objects with no blueshifted  $V_{circ} > 35 \kms$ DM halos in the
5 --7$ \mpc$  and 3 --  7 $\mpc$   shells. Figure \ref{fig8}
shows a scatter plot of \sigh\ against the mean density within the LV,
for the same objects considered in Figure \ref{fig7}.

As expected,  both Table \ref{table:blue} and Figure \ref{fig7} show that
the constraint of no having  blueshifted galaxies implies  LG-like objects with
cold environments. Both high-resolution simulations have yielded objects with no
blueshifted galaxies.   As can be deduced from   Table \ref{table:blue}, 
the closer to the LG we look for blueshifted galaxies, the less number
of LG's pass the criterion. Given the uncertainties involved in
the association of   $V_{circ}  \gtrsim  35 \kms$  halos with nearby faint
galaxies, we think that the number of LG-like objects with no
blueshifted galaxies cannot be used as a reliable statistic of the local
environment. Yet, it provides another strong observational evidence for
the coldness of the local flow. Figure \ref{fig8} provides further
support to our earlier findings about  the relation between \sigh\ and
the mean density of the LV (Figure \ref{fig4}). LG-like objects with no
blueshifted 'galaxies' in their LV are cold, $\sigma_H \lesssim 100 \kms$, and they
resides in LV with   $\rho_\mathrm{DM}/\bar{\rho}_\mathrm{DM} \lesssim  2$.

\begin{table*} 
  \begin{center}
    \begin{tabular}{c|cc|cc}
      \hline
      no negative velocity halos & \multicolumn{2}{c}{\HRLCDM} & \multicolumn{2}{c}{\LCDM160} \\ 
      \hline
              & LG candidates & $\sigma_H^{median}$  & LG candidates & $\sigma_H^{median}$ \\
      \hline
       all            & 12    &   97 $\pm$ 61   &    131     &  115 $\pm$ 58                \\
       5-7 Mpc  & 7      &   85 $\pm$ 15   &      46     &   90 $\pm$ 31                   \\
       3-7 Mpc  & 4      &   85 $\pm$ 16   &      15     &   95 $\pm$ 31                    \\
       2-7 Mpc  & 1      &   69 $\pm$ --     &        6     &   99 $\pm$ 19                     \\
 \hline 
 \end{tabular}
    \caption{The case of LG-like objects with no blueshifted galaxies:
      The left column presents the selection criteria of the LG-like objects:
      all - the benchmark \pairs, the following raws corresponds to
      LG-like objects  with no blueshifted DM halos with $V_{circ} > 35
      \kms$, taken as a proxy to observed galaxies,  within the
      respective shell. Each  raw provides the number of such groups and
      the median of \sigh\  and its  standard deviation  measured within their
      LV.  The analysis was  applied to the two high resolution
      simulations, \LCDMhr\  and \LCDM160.
 }
    \label{table:blue}
  \end{center}
\end{table*}

\section{Discussion}
\label{sec:disc}

A suite of CDM simulations have been used to study the nature of the
velocity field around Local Group - like objects. The objects are
selected so as to reproduce the main dynamical properties of the LG,
which include the range of masses of the two most massive DM halos in
the group, the proximity to a Virgo-like cluster and the absence of
nearby massive halos. The flow field has been analyzed by means of
\sigh, which is defined by the RMS value of the residual from the pure
Hubble flow, of all galaxies within   the Local Volume, defined to be
the volume enclosed within a radius of $7 \mpc$ around the LG. The
results are compared with the \sigh\ calculated for the CNG data. 
Two important conclusions are reached here. The first, and the more
important one, is  that there is no {\em 'coldness of the local flow'}
problem, but rather it is the relation between \sigh\  and the mean
local matter density that needs to be addressed in the context of the
LV. The other is that  within the canonical  
CDM  cosmological models  
$\sigma_H$ depends only on the mean  matter density and not 
on the cosmological constant. Recently, 
\cite{tikhonov09}  have  found a similar cold flow around
LG-like objects  in  a model  with Warm
Dark Matter (WDM) of 1 keV particle masses  which leads to a cut-off in the power
spectrum above $k_{\rm peak} = 3.7 \hmpc$.

The main result that has been found is that for all models considered
here there is a non-negligible fraction of the LG-like objects with
$\sigma_H$ close to the observed value. 
The two \LCDM\  simulations in the box $64 \hmpc$  have about
$(30  - 45)$\%, 
the \LCDM160\ has  25\%, the two OCDM simulations have roughly
$(50 - 65)$\%  
and even the SCDM model has 17$\%$, of their (\pairs) LG-like objects
colder  than the observed value.
However the scatter in $\sigma_H$ is not random but it correlates strongly
with the mean matter density within the LV, with higher $\sigma_H$ expected for
higher density.  
In the  canonical \LCDM\  case a $\sigma_H \lesssim 100 \kms$ implies that the mean
LV density is less than twice mean cosmological density. This is also
consistent with the requirement that there are no blueshifted galaxies
in the LV around the LG. It follows that
the old 'coldness of local flow'
problem reappears as a potentially new problem,  namely the local
density - coldness relation. The recent compilation of the abundance of
galaxies in the LV of \cite{tik08} provides interesting results.  These
authors studied the luminosity function of the galaxies in the CNG data
and estimated the  LV fractional density to be about $1.4\pm 0.17$
within $R=8 \mpc$. Using DM halos with $V_{circ} > 35 \kms$ as a proxy
to luminous galaxies, the \LCDM\ predicted density enhancement in the
LV  stands in    good agreement with the observed value.  
The realization that \sigh\ and the local density are closely connected
provides another reminder to the fact that 
progress in understanding the local dynamics depends on a parallel advance in understanding galaxy formation.

It has been speculated that the coldness of the local flow is a
manifestation of the dark energy, which dominates the mass-energy
density of the universe (\citeauthor{bar01} \citeyear{bar01},
\citeauthor{che04} \citeyear{che04}, \citeyear{che07},
\citeauthor{mac05}  \citeyear{mac05}). The basic argumentation behind
that claim is that very locally the gravitational field is supposed to
be dominated by the cosmological constant and hence  the induced Hubble flow is expected to be cold.
This claim has been refuted by \cite{hf08} who showed that within a
local volume defined by $0.75 \leq R \leq 2$ Mpc the flow is dominated by the
DM, regardless of the existence of a cosmological constant. Namely, the
OCDM and \LCDM\ models  yield virtually the same $\sigma_H$ within that
volume. 
The current paper extends the numerical experiments 
and reinforces the results of \cite{hf08}. 
That earlier work 
is extended to cover more simulations of higher resolution
and the LV is defined to extend over $1 \leq R \leq 7$ Mpc.

The current work substantiate and gives justification to the theoretical arguments suggested
by \cite{hf08} as to why dark energy does not dictates the local dynamics. The arguments 
on the role of the dark energy are all based on approximating the local dynamics by the 
monopole term, i.e. a gravitational field induced by an isolated point-like object. 
This would have implied a very tight correlation between \sigh\ and the mean overdensity 
within the LV, contrary to the large scatter manifested by  Figure \ref{fig7}. 
The arguments advocating the cooling of the local flow by the dark energy completely ignore the
environmental effects on the flow. These effects are very clearly demonstrated in Table \ref{tabrelax} 
which shows that \sigh\ nearly doubles by removing from the benchmark criteria  the constraint on the lack of MW-like halos within a sphere of $3\mpc$. 
The proximity of Virgo-like halos also changes the value of \sigh\ . 
It follows that the dynamics of quasi-linear objects such as the LG cannot be properly modeled without a detailed accounting 
of environmental effects.

The present paper reaches radically different conclusions than those
expressed in  \cite{mac05} concerning the role of the cosmological
constant in shaping the dynamics of the LV.
Yet, a close inspection shows that there is no technical disagreement
between the two papers.  We were careful to
follow the selection criteria of LG-like objects of \cite{mac05} and
hence their  \LCDM\ simulation can be directly compared with the  
present ones. In
  spite of the somewhat different value of the physical parameters of
  the \LCDM\ model used in the two papers the scatter of $\sigma_H$  
and its
  dependence of the local density are in close agreement.  A careful
  reading of \cite{mac05} reveals that  their statement concerning the
  role of the $\Lambda$ term is based on
  the comparison these authors did with the OCDM simulation of   
\cite{gov97}.

A careful comparison with the \sigh\ - $\delta \rho / \rho$ relation
of the LG candidates in the OCDM simulation shown  in Fig. 11 of
\cite{gov97},  and its presentation in Fig. 2 of  \citeauthor{mac05}
\citeyear{mac05},  reveals that it can be reproduced  in the present   
OCDM simulations
  by omitting   the Virgo constraints on the selection of the LG-like
  objects. Given the present  much better resolution, the number of  
our  LG  candidates is considerable larger than in
\cite{gov97}, in particular in the densest environments.
The linear fit shown in   Fig. 11 of \citeauthor{gov97} is biased  
towards the LGs located in $\delta  \rho / \rho  \le 1$, where   most  
of their LGs are found. Restricting  our fit to LGs with $\delta   
\rho / \rho  \le 1$, a quite similar fit emerges.
This indicates that the old and present  OCDM simulations equally  
reproduce, to within  their resolution limits, the same kind of LG  
candidates.
To summarize, by neglecting the Virgo constraints on the selection of  
LG-like objects and by restricting the analysis to less dense objects
the old  OCDM results are recovered.  This might hint   that the OCDM  
\sigh\ analysis of \cite{gov97}
was performed with respect to their 'cat2' catalog rather than the
stated 'cat3' catalog, in which the Virgo constraints were
imposed.

 No substantial differences are found between the constrained and unconstrained simulations. 
 The constraints imposed here are affecting the structure on scales
 larger than $\approx 5 \hmpc$ \citep{kly03}, a scale that coincides with the
 LV. This implies that the behavior of \sigh\ predominantly depends on
 the internal  
 dynamics within the LV and less on the one induced by larger scales. 
 This also  gives further support to the claim that, apart from the
 particularities of the very nearby structure, the universe around us
 constitutes a very typical realization of the  CDM cosmogonies.  
The particular   dynamical attributes of the LG are its mass range of
about $10^{12}\hmsun$, its binary nature,   the proximity to the Virgo
cluster and the absence of similar nearby objects. These have been used
as the selection criteria of LG-like objects in the simulations. In
\S\ref{sec:relax} these criteria have been relaxed, while keeping the
mass to be roughly the LG mass.  
One expects that the  proximity of the Virgo cluster heats the flow
field in the vicinity of the LG, due to its tidal field which induces
non-isotropic motions around the LG. Hence the relaxation of the
condition for a nearby Virgo-like cluster leads to  colder LG-like
objects. Similarly, in  LG-like objects with a single MW-like halo  the
gravitational field is dominated by a monopole term and this leads to a
colder flow than in binary LG-like systems. On the other hand the
largest, yet  opposite, effect is introduced by allowing MW-like
halos to be  close to the LG. The presence of such nearby halos heats 
the flow considerably. It follows that the main ingredient that is responsible
for the coldness of the local flow is the relative isolation, and
therefore also the low mean local density, of the LG. Furthermore, in
the CDM  cosmologies the coldness of the local flow depends more on the
properties of the LG  than on the possible presence of a dark energy
component.

\section{Acknowledgements}
Fruitful discussions 
and correspondence 
with  I. D. Karachentsev, A. Klypin, A. Maccio and A. Tikhonov are gratefully acknowledged.
This research has been supported by  the ISF (13/08 at the HU).
The support of the European Science Foundation through the ASTROSIM Exchange
Visits Program is acknowledged (SG, YH).  The simulations were
performed and analyzed on the Leibniz Rechenzentrum Munich (LRZ), the Barcelona
Supercomputing Center (BSC) and the Magerit supercomputer at the Centro
de Supercomputaci\'on y Visualizaci\'on de Madrid (CeSViMa).
 SG acknowledges a Schonbrunn Fellowship at the Hebrew University
 Jerusalem. GY would like to thank also MEC (Spain) for financial support
under project numbers FPA2006-01105 and AYA2006-15492-C03. 
LAMV acknowledges financial support from Comunidad de Madrid through a PhD fellowship.


\begin{thebibliography}{}



\bibitem[\protect\citeauthoryear{{Baryshev}, {Chernin} \&
  {Teerikorpi}}{{Baryshev} et~al.}{2001}]{bar01}
{Baryshev} Y.~V.,  {Chernin} A.~D.,    {Teerikorpi} P.,  2001, \aap, 378, 729


 \bibitem[{{Chernin} {et~al.}(2007){Chernin},
     {Karachentsev}, {Teerikorpi}, {Valtonen}, {Byrd}, {Efremov},
     {Dolgachev}, {Domozhilova}, {Makarov}, \& {Baryshev}}]{che07}
   {Chernin}, A.~D., {Karachentsev}, I.~D., {Teerikorpi}, P.,
   {Valtonen}, M.~J., {Byrd}, G.~G., {Efremov}, Y.~N., {Dolgachev},
   V.~P., {Domozhilova}, L.~M., {Makarov}, D.~I., \& {Baryshev}, Y.~V.
   2007, ArXiv e-prints, 0706.4068

 \bibitem[{{Chernin} {et~al.}(2004){Chernin}, {Karachentsev},
     {Valtonen}, {Dolgachev}, {Domozhilova}, \& {Makarov}}]{che04}
   {Chernin}, A.~D., {Karachentsev}, I.~D., {Valtonen}, M.~J.,
   {Dolgachev}, V.~P., {Domozhilova}, L.~M., \& {Makarov}, D.~I. 2004,
   \aap, 415, 19

\bibitem[\protect\citeauthoryear{{Davis} \& {Peebles}}{{Davis} \&
  {Peebles}}{1983}]{dav83}
{Davis} M.,  {Peebles} P.~J.~E.,  1983, \apj, 267, 465

\bibitem[\protect\citeauthoryear{{Freedman}, {Madore}, {Gibson}, {Ferrarese},
  {Kelson}, {Sakai}, {Mould}, {Kennicutt} Jr., {Ford}, {Graham}, {Huchra},
  {Hughes}, {Illingworth}, {Macri} \& {Stetson}}{{Freedman}
  et~al.}{2001}]{fre01}
{Freedman} W.~L.,  {Madore} B.~F.,  {Gibson} B.~K.,  {Ferrarese} L.,  {Kelson}
  D.~D.,  {Sakai} S.,  {Mould} J.~R.,  {Kennicutt} Jr. R.~C.,  {Ford} H.~C.,
  {Graham} J.~A.,  {Huchra} J.~P.,  {Hughes} S.~M.~G.,  {Illingworth} G.~D.,
  {Macri} L.~M.,    {Stetson} P.~B.,  2001, \apj, 553, 47

\bibitem[\protect\citeauthoryear{{Gill}, {Knebe} \& {Gibson}}{{Gill}
  et~al.}{2004}]{amiga}
{Gill} S.~P.~D.,  {Knebe} A.,    {Gibson} B.~K.,  2004, \mnras, 351, 399


\bibitem[\protect\citeauthoryear{{Gottl\"ober} \& {Klypin}}{{Gottl\"ober} \&
  {Klypin}}{2008}]{ART}
{Gottl\"ober} S.,  {Klypin} A.,  2008, in {{Wagner}, S. and {Steinmetz}, M.and
  {Bode}, A. and {Brehm}, M.} ed., {High Performance Computing in Science and
  Engineering,} {The ART of Cosmological Simulations}.
{Springer-Verlag}, p.~29


\bibitem[\protect\citeauthoryear{{Governato}, {Moore}, {Cen}, {Stadel}, {Lake}
  \& {Quinn}}{{Governato} et~al.}{1997}]{gov97}
{Governato} F.,  {Moore} B.,  {Cen} R.,  {Stadel} J.,  {Lake} G.,    {Quinn}
  T.,  1997, New Astronomy, 2, 91

\bibitem[\protect\citeauthoryear{{Hoffman}, {Martinez-Vaquero}, {Yepes} \&
  {Gottl{\"o}ber}}{{Hoffman} et~al.}{2008}]{hf08}
{Hoffman} Y.,  {Martinez-Vaquero} L.~A.,  {Yepes} G.,    {Gottl{\"o}ber} S.,
  2008, \mnras, 386, 390

\bibitem[\protect\citeauthoryear{{Karachentsev}, {Karachentseva}, {Huchtmeier}
  \& {Makarov}}{{Karachentsev} et~al.}{2004}]{kar04}
{Karachentsev} I.~D.,  {Karachentseva} V.~E.,  {Huchtmeier} W.~K.,    {Makarov}
  D.~I.,  2004, \aj, 127, 2031

\bibitem[\protect\citeauthoryear{{Karachentsev}, {Makarov}, {Sharina},
  {Dolphin}, {Grebel}, {Geisler}, {Guhathakurta}, {Hodge}, {Karachentseva},
  {Sarajedini} \& {Seitzer}}{{Karachentsev} et~al.}{2003}]{kar03}
{Karachentsev} I.~D.,  {Makarov} D.~I.,  {Sharina} M.~E.,  {Dolphin} A.~E.,
  {Grebel} E.~K.,  {Geisler} D.,  {Guhathakurta} P.,  {Hodge} P.~W.,
  {Karachentseva} V.~E.,  {Sarajedini} A.,    {Seitzer} P.,  2003, \aap, 398,
  479


\bibitem[\protect\citeauthoryear{{Karachentsev}, {Kashibadze}, {Makarov} \& {Tully}}
{{Karachentsev} et~al.}{2009}]{kar09}
{Karachentsev} I.~D., {Kashibadze} O.~G., {Makarov} D.~I., {Tully} R.~B. ,  2009, \mnras, 393, 1265




\bibitem[\protect\citeauthoryear{{Klypin}, {Gottl{\"o}ber}, {Kravtsov} \&
  {Khokhlov}}{{Klypin} et~al.}{1999}]{bdm}
{Klypin} A.,  {Gottl{\"o}ber} S.,  {Kravtsov} A.~V.,    {Khokhlov} A.~M.,
  1999, \apj, 516, 530

\bibitem[\protect\citeauthoryear{{Klypin}, {Hoffman}, {Kravtsov} \&
  {Gottl{\"o}ber}}{{Klypin} et~al.}{2003}]{kly03}
{Klypin} A.,  {Hoffman} Y.,  {Kravtsov} A.~V.,    {Gottl{\"o}ber} S.,  2003,
  \apj, 596, 19

\bibitem[\protect\citeauthoryear{{Kravtsov}, {Klypin} \& {Hoffman}}{{Kravtsov}
  et~al.}{2002}]{kra02}
{Kravtsov} A.~V.,  {Klypin} A.,    {Hoffman} Y.,  2002, \apj, 571, 563

\bibitem[\protect\citeauthoryear{{Macci{\`o}}, {Governato} \&
  {Horellou}}{{Macci{\`o}} et~al.}{2005}]{mac05}
{Macci{\`o}} A.~V.,  {Governato} F.,    {Horellou} C.,  2005, \mnras, 359, 941

\bibitem[\protect\citeauthoryear{{Martinez-Vaquero}, {Yepes} \&
  {Hoffman}}{{Martinez-Vaquero} et~al.}{2007}]{mv07}
{Martinez-Vaquero} L.~A.,  {Yepes} G.,    {Hoffman} Y.,  2007, \mnras, 378,
  1601

\bibitem[\protect\citeauthoryear{{Reiprich} \& {B{\"o}hringer}}{{Reiprich} \&
  {B{\"o}hringer}}{2002}]{rei02}
{Reiprich} T.~H.,  {B{\"o}hringer} H.,  2002, \apj, 567, 716

\bibitem[\protect\citeauthoryear{{Sandage} \& {Tammann}}{{Sandage} \&
  {Tammann}}{1975}]{san75}
{Sandage} A.,  {Tammann} G.~A.,  1975, \apj, 196, 313

\bibitem[\protect\citeauthoryear{{Sandage}, {Tammann} \& {Hardy}}{{Sandage}
  et~al.}{1972}]{san72}
{Sandage} A.,  {Tammann} G.~A.,    {Hardy} E.,  1972, \apj, 172, 253

\bibitem[\protect\citeauthoryear{{Schlegel}, {Davis}, {Summers} \& {Holtzman}}
{Schlegel et~al.}{1994}]{sch94} 
{Schlegel} D., {Davis} M., {Summers} F., {Holtzman} J.~A., 1994, \apj, 427, 527 



\bibitem[Spergel et al.(2007)]{wmap3} Spergel, D.~N., et al.\ 
2007, \apjs, 170, 377 

\bibitem[Spergel et al.(2003)]{wmap1} Spergel, D.~N., et al.\ 
2003, \apjs, 148, 175 



\bibitem[\protect\citeauthoryear{{Springel}}{{Springel}}{2005}]{gadget2}
{Springel} V.,  2005, \mnras, 364, 1105

\bibitem[\protect\citeauthoryear{{Tikhonov} \& {Karachentsev}}{{Tikhonov} \&
 {Karachentsev}}{2006}]{tik06}
{Tikhonov} A.,  {Karachentsev} I.D.,  2006, \apj, 653, 969


\bibitem[\protect\citeauthoryear{{Tikhonov} \& {Klypin}}{{Tikhonov} \&
  {Klypin}}{2008}]{tik08}
{Tikhonov} A.,  {Klypin} A.,  2008, ArXiv e-prints, 0807.0924


\bibitem[\protect\citeauthoryear{Tikhonov et al.}{2009}]{tikhonov09}
Tikhonov, A.V., Gottl\"ober, S., Yepes, G. Hoffman, Y.,  in preparation, (2009) 

\bibitem[\protect\citeauthoryear{{Tonry}, {Dressler}, {Blakeslee}, {Ajhar},
  {Fletcher}, {Luppino}, {Metzger} \& {Moore}}{{Tonry} et~al.}{2001}]{sbf}
{Tonry} J.~L.,  {Dressler} A.,  {Blakeslee} J.~P.,  {Ajhar} E.~A.,  {Fletcher}
  A.~B.,  {Luppino} G.~A.,  {Metzger} M.~R.,    {Moore} C.~B.,  2001, \apj,
  546, 681

\bibitem[\protect\citeauthoryear{{Tully}, {Shaya} \& {Pierce}}{{Tully}
  et~al.}{1992}]{tul92}
{Tully} R.~B.,  {Shaya} E.~J.,    {Pierce} M.~J.,  1992, \apjs, 80, 479

\bibitem[\protect\citeauthoryear{{Willick}, {Courteau}, {Faber}, {Burstein},
  {Dekel} \& {Strauss}}{{Willick} et~al.}{1997}]{mark}
{Willick} J.~A.,  {Courteau} S.,  {Faber} S.~M.,  {Burstein} D.,  {Dekel} A.,
   {Strauss} M.~A.,  1997, \apj, 109, 333

\bibitem[\protect\citeauthoryear{{Zaroubi}, {Hoffman} \& {Dekel}}{{Zaroubi}
  et~al.}{1999}]{hof99}
{Zaroubi} S.,  {Hoffman} Y.,    {Dekel} A.,  1999, \apj, 520, 413




\end{thebibliography}
\end{document}